\documentclass[12pt,dvips]{article}
\usepackage{rotating}
\usepackage{epsfig}
\usepackage{color}
\usepackage{cite}
\usepackage{amsmath}

\newcommand{\lsim}{\raisebox{-0.13cm}{~\shortstack{$<$ \\[-0.07cm] $\sim$}}~}

\begin{document}

\def\thefootnote{\fnsymbol{footnote}}

\begin{flushright}
{\tt  KIAS Preprint P12006}\\
January 2012
\end{flushright}

\begin{center}
{\bf {\Large
Singlino-driven Electroweak Baryogenesis \\[2mm]
in the Next-to-MSSM 
} }
\end{center}

\medskip

\begin{center}{\large
Kingman Cheung$^{a,b}$,
Tie-Jiun Hou$^{c}$,
Jae~Sik~Lee$^a$, and
Eibun Senaha$^d$ }
\end{center}

\begin{center}
{\em $^a$ Department of Physics, National Tsing Hua University, Hsinchu, Taiwan
300}\\[0.2cm]
{\em $^b$Division of Quantum Phases \& Devices,
Konkuk University,  Seoul 143-701, Korea}\\[0.2cm]
{\em $^c$Institute of Physics, Academia Sinica, Taipei, Taiwan 115}\\[0.2cm]
{\em $^d$School of Physics, KIAS, Seoul 130-722, Korea}
\end{center}

\bigskip\bigskip

\centerline{\bf ABSTRACT}

\noindent
We explore a new possibility of 
electroweak baryogenesis in the next-to-minimal supersymmetric standard model.
In this model, 
a strong first-order electroweak phase transition can be 
achieved due to the additional singlet Higgs field. 
The new impact of its superpartner (singlino) on the baryon asymmetry 
is investigated by employing the closed-time-path formalism.
We find that the $CP$ violating source term fueled by the singlino 
could be large enough to 
generate the observed baryon asymmetry of the Universe
without any conflicts with the current 
constraints from the non-observation of 
the Thallium, neutron and Mercury electric dipole moments.


\newpage


The origin of the baryon asymmetry of the Universe (BAU) is one of the longstanding problems
in particle physics and cosmology. 
The BAU is observed from the primordial abundances of the light elements 
(D, $^3$He, $^4$He, and $^7$Li) induced by the big-bang 
nucleosynthesis (BBN) and the cosmic microwave background (CMB) data.  
{}From these data the baryon-to-entropy ratio is found 
to be~\cite{Nakamura:2010zzi}
\begin{equation}
Y_B=\frac{n_{B}}{s}=\left\{
\begin{array}{ll}
(7.2 - 9.2) \times 10^{-11} & {\rm BBN} \\
(8.61 - 9.09) \times 10^{-11} & {\rm CMB} 
\end{array}
\right.\,.\label{YB_ob}
\end{equation}
The baryon asymmetry of the Universe must arise after the 
inflation if it happened, and before the BBN era.
As pointed out by Sakharov~\cite{Sakharov:1967dj}
the following three conditions must be satisfied in order to generate 
the BAU: (i) baryon-number violation, (ii) $C$ and $CP$ violation, and (iii) 
the departure from thermal equilibrium. 
The last condition could be evaded if the $CPT$ symmetry is violated.
Although the standard model (SM) of particle physics 
can in principle satisfy all the above three conditions, it turns out that
the $CP$ violation coming from the Cabbibo-Kobayashi-Maskawa matrix~\cite{CKM}
is too feeble to generate sufficient BAU~\cite{ewbg_sm_cp}.
Furthermore, the electroweak phase transition (EWPT) in the SM 
is a smooth crossover for a Higgs boson with a mass above the
LEP bound ($\sim 115$ GeV)~\cite{crossover}, and therefore renders the condition of 
departure from the thermal equilibrium infeasible. Thus, physics 
beyond the SM is indispensable to address this issue.

Supersymmetry (SUSY) is an attractive framework for new physics 
since it can provide elegant explanations for a number of key questions that
cannot be accommodated in the SM~\cite{SUSY}. 
Supersymmetry can stabilize the electroweak-symmetry breaking scale by 
providing a natural cancellation mechanism of the quadratic divergences
thus solving the gauge-hierarchy problem, offer a good candidate 
for the cold dark matter, enable the gauge-coupling unification, 
break electroweak symmetry dynamically,
and generate sufficient BAU.
The minimal realization of SUSY, known as the minimal supersymmetric
standard model (MSSM), is very attractive because of its simplicity.
It is, however, known to suffer from the so-called $\mu$ 
problem~\cite{mu-problem} 
and the little-hierarchy problem as noted more recently.
Moreover, the electroweak baryogenesis (EWBG) scenario~\cite{ewbg}
in the MSSM is highly restricted by 
current experimental data~\cite{recentMSSMEWBG,Funakubo:2009eg,Carena:2008vj}.
Especially, a Higgs boson which has escaped
the detection at LEP~2~\cite{Schael:2006cr}
and/or perhaps weighs around 125 GeV as hinted by 
the recent ATLAS/CMS results~\cite{LHC_SM_Higgs}
has pushed the MSSM to the edge of the allowed parameter space that 
can be consistent with a first-order EWPT~\cite{Carena:2008vj}.
This tension, however, could be relaxed if the MSSM is extended~\cite{EWBG_NMSSM,Funakubo:2005pu,EWBG_nMSSM,EWBG_UMSSM,EWBG_sMSSM,Kanemura:2011fy,Blum:2010by}.

In this Letter, we consider the 
next-to-minimal supersymmetric standard model (NMSSM)~\cite{NMSSM:review},
in which the $\mu$ problem is solved naturally.
It has been shown that a strong first-order EWPT can happen in 
the NMSSM much more easily
than in the MSSM~\cite{EWBG_NMSSM,Funakubo:2005pu}.
In the presence of the additional gauge-singlet field, 
the NMSSM Higgs potential contains
explicit/spontaneous $CP$ violation even at the tree 
level~\cite{Cheung:2010ba}.
Furthermore, this new additional source of $CP$ violation may give 
rise to sufficient BAU even when the MSSM fails.


%
The NMSSM superpotential is
\begin{equation}
  \label{Wpot}
W_{\rm NMSSM}\ =\ \widehat{U}^C {\bf h}_u \widehat{Q} \widehat{H}_u\:
+\:   \widehat{D}^C {\bf h}_d \widehat{H}_d \widehat{Q}  \: +\:
\widehat{E}^C {\bf h}_e \widehat{H}_d \widehat{L} \: +\:
\lambda \widehat{S} \widehat{H}_u \widehat{H}_d\ \: + \:
\frac{\kappa}{3}\ \widehat{S}^3 \ ,
\end{equation}
where $\widehat{S}$ denotes the singlet Higgs superfield,
$\widehat{H}_{u,d}$  are the two SU(2)$_L$ doublet Higgs superfields, and
$\widehat{Q}$,  $\widehat{L}$ and $\widehat{U}^C$,  $\widehat{D}^C$,
$\widehat{E}^C$ are the matter doublet and singlet superfields, respectively,
related to the up- and  down-type quarks and  charged leptons.
%
%
We note that the superpotential respects an extra discrete $Z_3$ symmetry.

The tree-level Higgs potential includes the terms coming from the
soft SUSY-breaking terms:
\begin{eqnarray}
V_{\rm soft}&=&m_1^2 H_d^\dagger H_d+m_2^2 H_u^\dagger H_u
        +m_S^2|S|^2
        +\left(\lambda A_{\lambda}S H_u H_d
        -\frac{1}{3}\kappa A_\kappa S^3+{\rm h.c.}\right)\,,
\end{eqnarray}
in which $\lambda$, $\kappa$, $A_\lambda$, and $A_\kappa$
may contain nontrivial $CP$ phases.
After the neutral components of the two Higgs doublets and
the singlet develop their vacuum-expectation-values (VEVs),
one may have three rephasing-invariant combinations  of the
$CP$-odd phases in the tree-level potential~\cite{Cheung:2010ba}:
\begin{equation}
\phi^\prime_\lambda - \phi^\prime_\kappa\,; \ \ \
\phi^\prime_\lambda + \phi_{A_\lambda}\,; \ \ \
\phi^\prime_\lambda + \phi_{A_\kappa}
\end{equation}
where
$\phi^\prime_\lambda \equiv \phi_\lambda+\theta+\varphi$
and $\phi^\prime_\kappa \equiv \phi_\kappa+3\varphi$ with
$\theta$ and $\varphi$ parameterizing the overall $CP$ phases of the doublet
$H_u$ and the singlet $S$, respectively.
It turns out that the latter two $CP$ phases are
fixed up to a two-fold ambiguity by the two $CP$-odd
tadpole conditions, while the difference 
$\phi^\prime_\lambda - \phi^\prime_\kappa$
remains as an independent physical $CP$ phase.
This tree-level $CP$ violation is a distinctive feature of the NMSSM compared to
the MSSM.

The stringent constraint on the tree-level $CP$ phase 
$\phi^\prime_\lambda - \phi^\prime_\kappa$ may come from
the non-observation  of  the electric dipole  moments (EDMs)  for 
Thallium~\cite{Regan:2002ta},
the neutron~\cite{Baker:2006ts}, and 
Mercury ($^{199}{\rm Hg}$)~\cite{Romalis:2000mg,Griffith:2009zz}.
The detailed study shows that the maximal $CP$ phase  
$\phi^\prime_\lambda - \phi^\prime_\kappa \sim 90^\circ$
could still be compatible with the current EDM constraints
taking account of the uncertainties in the calculations of
the Mercury EDM
when the sfermions of the first two generations are heavier
than about $300$ GeV~\cite{Cheung:2011wn}.

In this Letter, we address the question whether
sufficient BAU can be generated
by the $CP$ phase $\phi^\prime_\lambda - \phi^\prime_\kappa \sim \pm 90^\circ$
when the $CP$ phases appearing in all the other soft
SUSY-breaking terms are not large enough for the BAU
because of the tight EDM constraints with the
SUSY particles within the reach of the LHC. 
Explicitly, we are taking
$\sin(\phi^\prime_\lambda+\phi_{A_f})=
\sin(\phi^\prime_\lambda+\phi_i)=0$ with $\phi_{A_f}$ and 
$\phi_i$ denoting the $CP$ phases of the soft trilinear parameters $A_f$ and
the three gaugino mass parameters $M_{i=1,2,3}$, respectively.

The scenario we are considering has an intermediate value of $\tan\beta$ with
small $v_S \sim v(T=0)\simeq~246~{\rm GeV}$:
\begin{eqnarray}
&&
\tan\beta=5\,, \ \
v_S=200~{\rm GeV}\,,
\nonumber \\
&&
|\lambda|=0.81\,, \ \ |\kappa|=0.08\,; \ \
|A_\lambda|=575~{\rm GeV}\,, \ \ |A_\kappa|=110~{\rm GeV} \,; \ \
\nonumber \\
&&
\phi_\lambda^\prime-\phi_\kappa^\prime=\pm 90^\circ\,,  \ \
{\rm sign}\,[\cos(\phi^\prime_\kappa+\phi_{A_\kappa})] =
{\rm sign}\,[\cos(\phi^\prime_\lambda+\phi_{A_\lambda})]  = + 1 \,.
\label{eq:scenario}
\end{eqnarray}
The other parameters are chosen as
\begin{eqnarray}
M_{{\widetilde Q}_{3}}=1~{\rm TeV}\,, M_{{\widetilde U}_{3}}=150~{\rm GeV} - 1~{\rm TeV}\,,
M_{{\widetilde D}_{3}}=250~{\rm GeV}  -  1~{\rm TeV}\,; \ \ 
|A_t|=|A_b|=1~{\rm TeV}\,.
\end{eqnarray}
We find $M_A\simeq 600~{\rm GeV}$ for the parameters chosen
and have taken $M_1=M_2=-200$ GeV to fix the neutralino 
and chargino sectors.
We find that a first-order phase transition could occur in some
regions of the parameter space of this scenario
which is needed for the EWBG.
Actually, the diverse patterns of the EWPT in the NMSSM have been investigated
in Ref.~\cite{Funakubo:2005pu}.
Among the patterns the so-called type-B transition opens a new possibility for 
the strong first-order EWPT. In this type of transitions, the lighter 
stop could be heavier than the top quark, in contrast to the MSSM
EWPT. Instead, the singlet Higgs field plays an essential role. 
In a typical parameter-space point, during the EWPT the VEVs 
$(v, v_S)$ change from $(0, 600~{\rm GeV})$ to 
$ (208~{\rm GeV}, 249~{\rm GeV})$ at the critical temperature $T_C=110$ GeV.
The dramatic change of $v_S$ is the most important
feature of the type-B transition. 
As discussed in Ref.~\cite{Funakubo:2005pu}, the relatively 
small $\kappa$ is required to ensure
the local minimum in the $v_S$ direction as in the scenario
Eq.~(\ref{eq:scenario}).

For the calculation of the baryon density, 
the Closed-Time-Path (CTP) formalism is employed~\cite{ctp,Lee:2004we} and
we closely follow Refs.~\cite{Lee:2004we,Huet:1995sh}
\footnote{For the more precise calculation, the profiles of the bubble wall
and the profile-dependent masses and widths 
should be taken into account when solving the 
quantum transport equations during the EWPT.}.
The baryon density in the broken phase is given by
\begin{eqnarray}
n_{B}&=&\frac{n_F\Gamma_{ws}}{2}\,{\cal A}\,
\left[ r_1 + r_2
\frac{v_w^2}{\Gamma_{ss}\bar{D}}\left(1-\frac{D_q}{\bar{D}}\right) \right]\,
\nonumber \\
&& 
\times
\frac{2\bar{D}D_q}{v_w\left[\bar{D}v_w+(2D_q-\bar{D})\sqrt{v_w^2+4{\cal R}D_q}\right]
+4{\cal R}\bar{D}D_q}  \;,
\end{eqnarray}
where $n_F=3$ is the number of fermion families,
$\Gamma_{ws} = 6\, \kappa\, \alpha_2^5\, T \sim 0.5 \times 10^{-5}\,T$
with $\kappa \simeq 20$ and $\alpha_2 \simeq 1/30$,
and
$\Gamma_{ss}=16 \kappa' \alpha_s^4 T$ with $\kappa' = {\cal O}(1)$.
We are taking the bubble wall velocity $v_w=0.04$.
The relaxation term 
${\cal R} = \Gamma_{ws}\,\left[\frac{9}{4}(1+{n_{sq}}/{6})^{-1}
+\frac{3}{2}\right]$ with $n_{sq}$ for the number of light squark flavors.
%
The diffusion constant is given by
\begin{eqnarray}
\bar{D}  = 
\frac{(9k_Qk_T+k_Bk_Q+4k_Tk_B)D_q+k_H(9k_T+9k_Q+k_B)D_h }
{9k_Qk_T+k_Bk_Q+4k_Tk_B+k_H(9k_T+9k_Q+k_B)}\,,
\end{eqnarray}
with $D_q=6/T$ and $D_h=110/T$~\cite{Joyce:1994zn}.
The $k$ factors are given by the sum
$k_{Q,T,B}=k_{q_L,t_R,b_R}+k_{\tilde{q}_L,\tilde{t}_R,\tilde{b}_R}$
and $k_H=k_{H_d}+k_{H_u}+k_{\tilde{H}}$ where 
\begin{equation}
k_i = g_i\,\frac{6}{\pi^2}\,\int_{m_i/T}^\infty
dx\,\frac{e^x}{(e^x\pm 1)^2}\,x\sqrt{x^2-m_i^2/T^2}\,.
\end{equation}
Note $g_i=1$ for a chiral fermion and a complex scalar.
For example,
$g_i(t_L)=g_i(\widetilde{t}_L)=3$ taking account of the 3 colors,
$g_i(H)=2$ for a Higgs doublet, and
$g_i(\widetilde{H})=2$ for a Dirac Higgsino.
We further note that $k_i=g_i$ and
$k_i=2g_i$  for a chiral fermion and a complex scalar, respectively,
in the zero-mass limit.
The coefficients $r_{1,2}$ are given by
\begin{eqnarray}
r_1  = \frac{9k_Qk_T-5k_Qk_B-8k_Tk_B}{k_H(9k_Q+9k_T+k_B)}\,, \ \ \
r_2  = \frac{k_B^2(5k_Q+4k_T)(k_Q+2k_T)}{k_H(9k_Q+9k_T+k_B)^2}\,.
\end{eqnarray}
Note that, in the limit of very heavy stops and sbottoms, the sfermion
contributions to their corresponding $k$ factors become negligible and 
one may have $k_Q=6$ and $k_T=k_B=3$, which leads
to vanishing $r_1$ to which the baryon density 
$n_B$ is directly proportional.
We also observe that the coefficient $r_1$ vanishes when the stops and
sbottoms are all degenerate.
In the calculation of the $k$ factors, we have taken 
the thermal masses given in Ref.~\cite{Chung:2009cb}
with appropriate modifications.
Finally, the parameter ${\cal A}$ is given by
\begin{equation}
{\cal A} \simeq 
k_H L_w\, \sqrt{\frac{r_\Gamma}{\bar{D}}} \ \
\frac{S_{\widetilde{S}\widetilde{H}^0}^{\rm CPV}}
{\sqrt{\Gamma_M^-+\Gamma_h}} \;,
\label{eq:cal_a}
\end{equation}
where $L_w$ denotes the bubble wall width
and
$r_\Gamma = \bar\Gamma / (\Gamma_M^-+\Gamma_h)$. 
The $CP$-conserving particle number changing rate is
\begin{eqnarray}
\bar{\Gamma}  = 
\frac{(9k_Q+9k_T+k_B)(\Gamma_M^-+\Gamma_h)}
{9k_Qk_T+k_Bk_Q+4k_Tk_B+k_H(9k_T+9k_Q+k_B)}\,,
\end{eqnarray}
where
\begin{equation}
\Gamma_M^- = \frac{6}{T^2}\,\left(\Gamma_t^- +
\Gamma_{\widetilde{t}}^- \right) \,, \ \ \
\Gamma_h = \frac{6}{T^2}\,\left(
\Gamma_{\widetilde{H}^\pm\widetilde{W}^\pm} +
\Gamma_{\widetilde{H}^0\widetilde{W}^0} +
\Gamma_{\widetilde{H}^0\widetilde{B}^0} +
\Gamma_{\widetilde{H}^0\widetilde{S}}
\right) \,, \ \ \
\end{equation}
with
$\Gamma_{\widetilde{H}\widetilde{X}}=\Gamma_{\widetilde{H}\widetilde{X}}^-
-\Gamma_{\widetilde{H}\widetilde{X}}^+$.
We refer to Ref.~\cite{Lee:2004we} for details of the
calculations of the rates
$\Gamma_{t\,,\widetilde{t}}^-$ and
$\Gamma^\pm_{\widetilde{H}^\pm\widetilde{W}^\pm\,,
\widetilde{H}^0\widetilde{W}^0\,,\widetilde{H}^0\widetilde{B}^0\,,\widetilde{H}^0\widetilde{S}}$.

Here we present the analytic expression for the
singlino-driven $CP$-violating source term
appearing in the parameter ${\cal A}$ in Eq.~(\ref{eq:cal_a})
as follows:
\begin{equation}
S_{\widetilde{S}\widetilde{H}^0}^{\rm CPV} 
= -2|\lambda|^2|M_{\widetilde{S}}||\mu_{\rm eff}|v^2
	\dot{\beta}\sin(\phi_\lambda-\phi_\kappa)\,
	\mathcal{I}_{\widetilde{S}\widetilde{H}^0}^f \;,
\end{equation}
where $|\mu_{\rm eff}|=|\lambda|v_S/\sqrt{2}$ and 
\begin{equation}
|M_{\tilde S} (T)| =
\left[ 2 |\kappa|^2 v_S^2 +
\frac{|\lambda|^2+2|\kappa|^2}{8}\,T^2
\right]^{1/2}
\end{equation}
including the singlino thermal mass term. 
We assume that there is no spontaneous $CP$ violation or $\theta=\varphi=0$.
We note that the source term vanishes when
$\sin(\phi_\lambda - \phi_\kappa)=0$ or
$\dot{\beta}=0$. 
If $\beta$ has a kink-type profile, $\dot{\beta}\simeq v_w \Delta\beta/L_w$
and ${\cal A}$ becomes independent of $L_w$.
In the MSSM, $\Delta\beta$ was found to 
be $\mathcal{O}(10^{-2}-10^{-3})$~\cite{Moreno:1998bq,Funakubo:2009eg}. 
We are taking $\Delta\beta=0.02$ in our estimation of the source term
\footnote{
Note the source term grows linearly with $\Delta\beta$ and our
choice is optimal.
We observe that the variation depending on the choice of $\Delta\beta$
is to be regarded as the theoretical uncertainty
and the precise determination of $\Delta\beta$ in the NMSSM is  
beyond the scope of this Letter.  }.
The fermionic source function ${\cal I}^f$ takes
the generic form of
\begin{align}
\mathcal{I}_{ij}^f &= \frac{1}{4\pi^2}\int_0^\infty dk~\frac{k^2}
	{\omega_i\omega_j}
\Big[
	\big(1-2{\rm Re}(n_j)\big)
	I(\omega_i, \Gamma_i,\omega_j,\Gamma_j)
	+\big(1-2{\rm Re}(n_i)\big)\,
        I(\omega_j, \Gamma_j,\omega_i,\Gamma_i)\nonumber\\
&\hspace{4cm}
	-2\big({\rm Im}(n_i)+{\rm Im}(n_j)\big)\,
	G(\omega_i,\Gamma_i,\omega_j, \Gamma_j)
\Big],
\end{align}
where
\begin{eqnarray}
n_{i}\equiv \frac{1}{e^{(\omega_{i}-i\Gamma_{i})/T}+1}
\end{eqnarray}
with $\omega_i=\sqrt{k^2+m^2_i}$.  
We note that the thermal width $\Gamma_i$ at finite temperature is given by
the imaginary part of its self-energy
which is nonvanishing independently of whether the particle
$i$ is stable or not~\cite{Weldon:1983jn}. Specifically, 
for the calculation of the source function,
we have taken the thermal widths given in Ref.~\cite{Chung:2009qs}
\footnote{For the thermal width of the singlino, we are taking
$\Gamma_{\widetilde{S}}=0.03\,T$ considering the large
coupling $|\lambda|=0.81$. 
We find $Y_B$ is affected by the amount of about (25-35) \%
as $\Gamma_{\widetilde{S}}$ varies between $0.003\,T$ and $0.03\,T$.}.
The thermal functions $I$ and $G$ are defined by
\begin{align}
I(a,b,c,d) 
&= \frac{1}{2}\frac{1}{[(a+c)^2+(b+d)^2]}\sin\left[2\arctan\frac{a+c}{b+d}\right] \nonumber\\
&	+\frac{1}{2}\frac{1}{[(a-c)^2+(b+d)^2]}\sin\left[2\arctan\frac{a-c}{b+d}\right], 
\\[0.2cm]
G(a,b,c,d)
&= -\frac{1}{2}\frac{1}{[(a+c)^2+(b+d)^2]}\cos\left[2\arctan\frac{a+c}{b+d}\right] \nonumber\\
&	+\frac{1}{2}\frac{1}{[(a-c)^2+(b+d)^2]}\cos\left[2\arctan\frac{a-c}{b+d}\right].
\end{align}
We note the thermal functions lead to
${\cal I}^f_{ij} \propto \Gamma_i + \Gamma_j$ when $m_i\sim m_j\,,T \gg \Gamma_i$.

\begin{figure}[!t]
\begin{center}
{\epsfig{figure=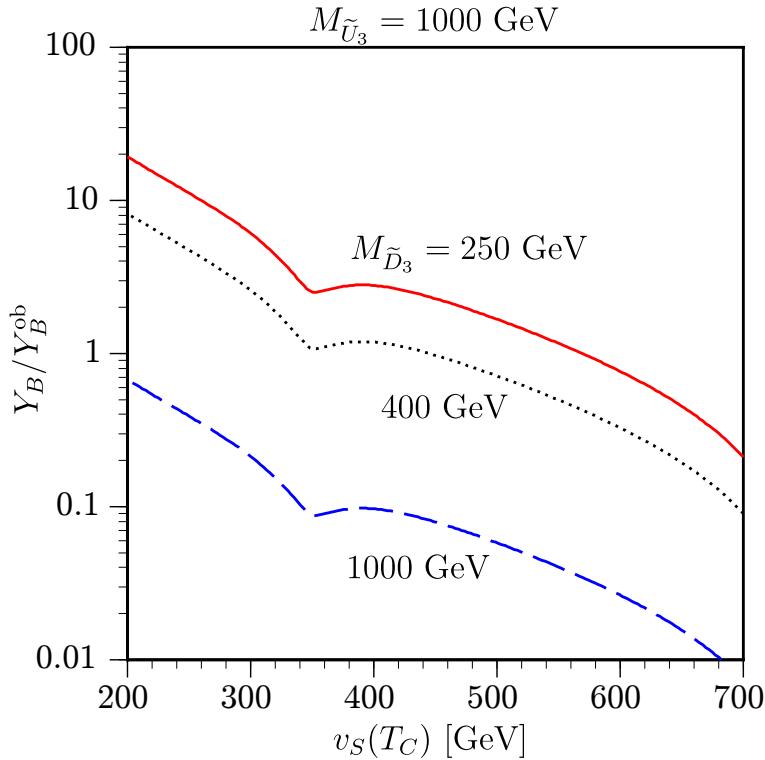,height=10.0cm,width=10.0cm}}
\end{center}
\caption{\it The singlino-driven $Y_B$ as functions of
the singlino VEV at the phase transition which, in principle,
can take any value between 250 GeV and 600 GeV in the type-B 
EWPT. We are normalizing our predictions to  $Y_B^{\rm ob}=8.8\times 10^{-11}$
for three values of $M_{{\widetilde D}_{3}}=250$ GeV,
$400$ GeV, and $1$ TeV.  
We fix $M_{{\widetilde Q}_{3}}=M_{{\widetilde U}_{3}}=1$ TeV.}
\label{fig:ybvsc}
\end{figure}
\begin{figure}[!t]
\begin{center}
{\epsfig{figure=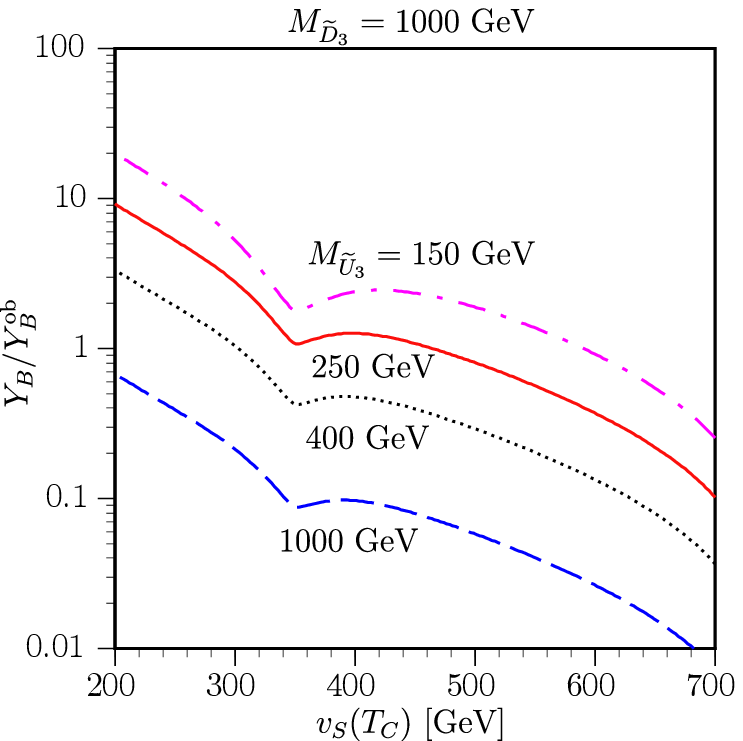,height=10.0cm,width=10.0cm}}
\end{center}
\caption{\it The same as in Fig.~\ref{fig:ybvsc} but varying
$M_{{\widetilde U}_{3}}$:
$M_{{\widetilde U}_{3}}=150$ GeV, $250$ GeV, $400$ GeV and $1$ TeV.
We fix $M_{{\widetilde Q}_{3}}=M_{{\widetilde D}_{3}}=1$ TeV.}
\label{fig:ybvsc_MstR}
\end{figure}
In Fig.~\ref{fig:ybvsc}, we show our predictions for the
singlino-driven  $Y_B/Y_B^{\rm ob}$ as functions of $v_S(T_C)$
taking three values of $M_{{\widetilde D}_{3}}=250$~GeV,
$400$~GeV, and $1$~TeV
with $Y_B^{\rm ob}$ being the averaged value of the two BAU data in Eq.~(\ref{YB_ob}).
Here $v_S(T_C)$ denotes the singlino VEV at $T_C$
and it can take on any value between 250 GeV and 600 GeV in the type-B EWPT:
\begin{equation}
(v, v_S)=(0, 600~{\rm GeV})\to(208~{\rm GeV}, 249~{\rm GeV})
\end{equation}
at $T_C=110$ GeV~\cite{Funakubo:2005pu}.  
We find that
the singlino-Higgsino mass difference becomes smaller
as $v_S$ decreases, leading to the larger $Y_B$. 
The more accurate determination of $Y_B$ requires
the knowledge of the profiles of the bubble wall
and the treatment of the diffusion equation beyond the 
formalism developed in Refs.~\cite{Huet:1995sh,Lee:2004we}.
We leave the more precise determination
of $Y_B$ in the NMSSM framework for future work~\cite{future1}. 

{}From Fig.~\ref{fig:ybvsc}, we can see that $Y_B$ is much suppressed 
when $M_{{\widetilde D}_{3}}=1$~TeV 
since the ratio $r_1$ is almost vanishing
when both the stops and sbottoms are heavy and/or degenerate.
However, $Y_B$ grows quickly as $M_{{\widetilde D}_{3}}$
decreases. When 
$M_{{\widetilde D}_{3}} = 400$~GeV, the ratio
$Y_B/Y_B^{\rm ob}$ is larger than $1$ in the region
$v_S(T_C)\lsim 440$ GeV. Furthermore, in the case with 
$M_{{\widetilde D}_{3}} = 250$~GeV,
sufficient BAU can be generated 
via the singlino-driven mechanism,
irrespective of the nonlinear dynamics during the EWPT.
We found the similar behavior by fixing $M_{\tilde{Q}_3}$ and $M_{\tilde{D}_3}$
and varying $M_{\tilde{U}_3}$, as shown in Fig. 2. We observe the same mass
$M_{\tilde{U}_3}$ would give a slightly smaller $Y_B$ compared to the same
mass of $M_{\tilde{D}_3}$.  Summarizing these results, we conclude that
a sizable mass splitting either in the stop 
sector or the sbottom sector is needed
for the successful singlino-driven BAU.
In passing, we note that the resonance enhancement of the widths 
$\Gamma_{\widetilde{H}^\pm\widetilde{W}^\pm\,,
\widetilde{H}^0\widetilde{W}^0\,,\widetilde{H}^0\widetilde{B}^0}$
induces the dip around $v_S(T_C)=350$ GeV 
where $|M_1|=|M_2|=|\mu_{\rm eff}|$.

Lastly,  the EWBG scenario considered here 
includes the two light Higgs states well below 100 GeV
escaping LEP constraints
\cite{Cheung:2010ba}
and the lightest neutralino of about 45 GeV with the
singlino fraction of $\sim$ 40 \%.
The light Higgs bosons and the lightest
neutralino deserve further studies
in connection with the current LHC Higgs searches and
the abundance of dark matter in the Universe

In this Letter, we have 
examined a new possibility of 
a singlino-driven mechanism 
for the BAU in the NMSSM framework.
In contrast to the MSSM,
explicit and/or spontaneous $CP$ violation can occur 
in the NMSSM Higgs potential even at the tree level.
We emphasize that this new source of 
$CP$ violation may solely give rise to the sufficient BAU
without any conflicts with the current EDM constraints,
as long as there is a sizable mass splitting in the stop and/or
the sbottom sectors
with the lighter stop and/or sbottom weighing 
below $\sim 500$ GeV~\cite{future1}.

\vspace{-0.2cm}
\subsection*{Acknowledgements}
\vspace{-0.3cm}
\noindent
E.S. would like to thank K. Funakubo, Y. Okada, Y.-F. Zhou and M. Asano 
for useful discussions on the CTP formalism.
The work was supported in parts by the National Science Council of
Taiwan under Grant Nos. 100-2112-M-007-023-MY3, 99-2112-M-007-005-MY3
and NSC-100-2811-M-001-037, and by the WCU program through the KOSEF
funded by the MEST (R31-2008-000-10057-0).

%
%
%


\end{document}